\documentclass[pra,11pt,amsmath,amssymb,superscriptaddress,eqsecnum]{revtex4}

\usepackage{bm,hyperref,ifpdf}
\usepackage[utf8]{inputenc} 
\ifpdf
\usepackage[pdftex]{graphicx}
\else
\usepackage{graphicx}
\fi

\begin{document}

\ifpdf
\DeclareGraphicsExtensions{.pdf, .jpg, .tif}
\else
\DeclareGraphicsExtensions{.pdf, .jpg}
\fi

\title{Nonlinear interferometry with Bose-Einstein condensates}

\author{Alexandre B.~Tacla}
\affiliation{Center for Quantum Information and Control,
MSC~07--4220, University of New Mexico, Albuquerque, New
Mexico~87131-0001, USA}

\author{Sergio Boixo}
\affiliation{Institute for Quantum Information, California
Institute of Technology,\\ Pasadena, California~91125, USA}

\author{Animesh Datta}
\affiliation{Clarendon Laboratory, Department of Physics,
University of Oxford, OX1 3PU, United Kingdom}

\author{Anil Shaji}
\affiliation{School of Physics, Indian Institute of Science
Education and Research, CET Campus, Thiruvananthapuram,
Kerala~695016, India}

\author{Carlton M.~Caves}
\affiliation{Center for Quantum Information and Control,
MSC~07--4220, University of New Mexico, Albuquerque, New
Mexico~87131-0001, USA}
\affiliation{School of Mathematics and
Physics, University of Queensland, Brisbane, Queensland 4072,
Australia}

\begin{abstract}
We analyze a proposed experiment [Boixo {\it et al.},\ Phys.\ Rev.\ Lett.\ {\bf 101}, 040403 (2008)] for achieving sensitivity scaling better than $1/N$ in a nonlinear Ramsey interferometer that uses a two-mode Bose-Einstein condensate (BEC) of $N$ atoms. We present numerical simulations that confirm the analytical predictions for the effect of the spreading of the BEC ground-state wave function on the ideal $1/N^{3/2}$ scaling. Numerical integration of the coupled, time-dependent, two-mode Gross-Pitaevskii equations allows us to study the several simplifying assumptions made in the initial analytic study of the proposal and to explore when they can be justified. In particular, we find that the two modes share the same spatial wave function for a length of time that is sufficient to run the metrology scheme.
\end{abstract}

\maketitle

\section{Introduction}

Recent advances in experimental techniques are providing access to
unprecedented levels of control over quantum systems and turning the
quest for the fundamental limits of metrology into a question of
practical importance, instead of just a theoretical curiosity.  The
success of many experiments that rely on weak-signal detection
inevitably depends on improvement of metrological methods that
operate near the limits established by quantum mechanics.

In this regard, several strategies have been proposed over the past
few years in order to make quantum-limited metrology accessible to
current experiments.  Most of these protocols focus on schemes for preparing optimal states, such as squeezed states, cat states, and path entangled states ($N00N$ states), to be fed into linear interferometers.  Ideally, these states achieve
sensitivities at the quantum limit for linear interferometry, often
referred to as the \textit{Heisenberg limit\/}~\cite{giovannetti06a}.
The optimal input states, however, are difficult to prepare and
very vulnerable to decoherence, thus making these protocols a major challenge
for experimental realization.  An alternative approach, using
nonlinear interferometry, has emerged as a promising way to
outperform $1/N$-limited linear interferometry without relying on
entanglement or
squeezing~\cite{luis04a,boixo08a,boixo08b,woolley08a,boixo09}.

In single-parameter estimation, the Heisenberg limit corresponds to
the best possible scaling of sensitivity with the resources available
for the measurement, here taken to be the number of quantum subsystems
available for the task.  This scaling is not universal, but rather
depends on the nature of the coupling between the quantum subsystems and
the parameter to be estimated, and is enhanced by nonlinear
couplings~\cite{luis04a,boixo07a,boixo08a,rey07,choi08}. Moreover, further
analysis showed that this enhancement is purely a dynamical effect,
which is independent of entanglement
generation~\cite{boixo08b,boixo09}.  This, in turn, makes the
enhanced scaling potentially robust against decoherence, as opposed to the
strategies previously mentioned.  Initial entangled states are
required to achieve the optimal scaling in nonlinear metrology, but
protocols that only involve separable states are sufficient to beat
the $1/N$ Heisenberg scaling of linear
metrology~\cite{luis04a,boixo08a,boixo08b,boixo09,woolley08a}.

In this paper, we analyze the experiment proposed
in~\cite{boixo08b,boixo09}, which uses a two-mode Bose-Einstein
condensate (BEC) of $N$ atoms to implement a nonlinear Ramsey
interferometer whose detection uncertainty scales better than the
optimal $1/N$ Heisenberg scaling of linear interferometry.  This
protocol takes advantage of the pairwise scattering interaction in a
BEC, which essentially allows interferometric phases to accumulate
$N$ times faster than in a linear interferometer, thus permitting a
measurement sensitivity that scales as $1/N^{3/2}$.

As was investigated analytically in previous work~\cite{boixo09},
there are various challenges and potential problems in implementing
such a measurement protocol and achieving the desired scaling.  In
view of currently available techniques and realistic experimental
parameters, we further investigate such issues by means of numerical
simulations and more accurate approximation procedures.

We first review, in Sec.~\ref{nonlinearinterferometry}, an
approximate analytical description of the proposed protocol, which
was presented in~\cite{boixo09}; we highlight the several simplifying
assumptions of this approximation.  We then compare, in
Secs.~\ref{numerics} and~\ref{improvedmodel}, these analytical
estimates and predictions with the results of numerical simulations.
By solving the time-independent Gross-Pitaevskii (GP) equation, we
first analyze, in Sec.~\ref{spreading}, how the spreading of the BEC
wave function with increasing $N$ affects the scaling and how this
effect can be suppressed by the use of low-dimensional, hard-trap
geometries.  Numerical integration of the time-dependent, two-mode GP
equations is presented in Sec.~\ref{dynamics}; these numerical
results simulate the fringe signal of the protocol, allowing us to
investigate how position-dependent phase shifts across the condensate
degrade the fringe visibility.  Section~\ref{improvedmodel} considers
further the differentiation of the spatial wave functions of the two
modes by investigating in some detail an alternative analytical
approximation to the two-mode dynamics, proposed in~\cite{boixo09},
which takes into account the effects of the position-dependent phase
shifts for times before the two modes separate spatially.
Section~\ref{conclusion} concludes with additional perspective on our
results.

\section{Nonlinear interferometry using BECs}
\label{nonlinearinterferometry}

As pointed out in~\cite{boixo08b,boixo09}, a two-mode BEC of $N$
atoms can be used to implement a nonlinear Ramsey interferometer
whose detection sensitivity scales better than $1/N$.  We briefly
review in this section key aspects of this metrology protocol and of
the zero-order model used to describe the evolution of the two-mode
BEC.  Further details can be found in~\cite{boixo09}.

\subsection{Model}

We consider a BEC of $N\gg 1$ atoms that can occupy two hyperfine
states, henceforth labeled $|1\rangle$ and $|2\rangle$. We assume the
BEC is at zero temperature and that all the atoms are initially
condensed in state $|1\rangle$ with wave function $\psi_N(\vec r)$,
which is the $N$-dependent solution (normalized to unity) of the
time-independent GP equation
\begin{equation}
\label{timeindepGPE}
\bigg(-\frac{\hbar^2}{2 m }\nabla^2
+V({\vec r})
+g_{11}N|\psi_N(\vec r)|^2 \bigg)\psi_N(\vec r)=\mu_N\psi_N(\vec r),
\end{equation}
where $V(\vec r)$ is the trapping potential, $\mu_N$ is the
chemical potential, and $g_{11}$ is the intraspecies scattering
coefficient.  This coefficient is determined by the $s$-wave
scattering length $a_{11}$ and the atomic mass $m$ according to
the formula $g_{11}=4\pi\hbar^2a_{11}/m$.

We describe the system by the so-called Josephson approximation, which assumes that both modes
have and retain the same spatial wave function $\psi_N(\vec r)$ from
Eq.~(\ref{timeindepGPE}).  In this approximation, the BEC dynamics is
governed by the two-mode Hamiltonian
\begin{equation}
\label{BEChamil1}
\hat H = N E_0 +
\frac{1}{2}\eta_N\!\sum_{\alpha,\beta=1}^2
g_{\alpha\beta}
\hat{a}_\beta^{\dagger}\hat{a}_\alpha^{\dagger}
\hat{a}_\alpha^{\vphantom{\dagger}}\hat{a}_\beta^{\vphantom{\dagger}}\;.
\end{equation}
Here $\hat{a}^\dagger_{\alpha}$ ($\hat{a}_{\alpha}$) creates
(annihilates) an atom in the hyperfine state $|\alpha\rangle$, with
wave function $\psi_N$,
$g_{\alpha\beta}=4\pi\hbar^2a_{\alpha\beta}/m$, $E_0$ is the
mean-field single-particle energy, given by
\begin{equation}
\label{eq:E0}
E_0=\int d^3 r
\left(\frac{\hbar^2}{2 m }|\nabla \psi_{N}|^2 + V({\vec r}) |\psi_{N}|^2\right),
\end{equation}
and the quantity
\begin{equation}
    \label{eta}
    \eta_N = \int d^3 r\,|\psi_{N}(\vec r)|^4
\end{equation}
is a measure of the inverse volume occupied by the condensate wave
function $\psi_N$. Notice that this effective volume renormalizes the
scattering interactions, thereby defining effective nonlinear
coupling strengths $g_{\alpha\beta}\eta_N$. The Josephson
approximation applies if one can drive fast transitions between the two
hyperfine levels, the two levels are trapped by the same
external potential, the atoms only undergo elastic collisions,
and the spatial dynamics are slow compared to the accumulation of
phases in the two hyperfine levels.  In addition, notice that the zero-temperature mean-field treatment of the Josephson Hamiltonian (\ref{BEChamil1}) assumes that the quantum depletion of the condensate is negligible. We make this assumption throughout on the grounds that the depletion is expected to be very small \cite{dalfovo99}.

The Josephson-approximation evolution is described in a more
convenient way in terms of Schwinger angular-momentum
operators~\cite{leggett01a}. Introducing the operator $\hat{J}_z =
(\hat{a}^\dagger_1\hat{a}_1-\hat{a}^\dagger_2\hat{a}_2)/2$, one finds
that Eq.~(\ref{BEChamil1}) can be written as
\begin{equation}
\label{BEChamil2}
\hat H = \gamma_1\eta_N N\hat{J}_z+
\gamma_2\eta_N \hat{J}_z^{\,2},
\end{equation}
where we define two new coupling constants that
characterize the interaction of the two modes,
\begin{equation}
\gamma_1 = \frac{1}{2}(g_{11}-g_{22}) \quad {\mbox{and}} \quad
\gamma_2 = \frac{1}{2}(g_{11}+g_{22})-g_{12}\;.
\end{equation}
We omit \emph{c-number} terms whose only effect is to introduce an
overall global phase.

The dynamics governed by Eq.~(\ref{BEChamil2}) is analogous to that
of an
interferometer with nonlinear phase shifters~\cite{boixo08b}.  Due to
the different scattering interactions, the first term of
Eq.~(\ref{BEChamil2}) introduces a relative phase shift that is
proportional to the total number of atoms in the condensate, whereas
the $\hat{J}_z^{\,2}$ term leads to more complicated dynamics that
create entanglement and phase diffusion.  Both terms can be used to
implement nonlinear metrology protocols whose phase detection
sensitivity scales better than $1/N$.  For initial product states,
the entanglement created by $\hat{J}_z^{\,2}$ has no influence on the
enhanced scaling and therefore offers no advantage over the $N
\hat{J}_z$ evolution.  On the contrary, it is better to avoid the associated phase
dispersion~\cite{boixo08a}, which can be accomplished
by a suitable choice of the condensate atomic species.

We consider a condensate of $^{87}$Rb atoms constrained to the
hyperfine levels $|F=1,M_F=-1\rangle \equiv|1\rangle$ and
$|F=2,M_F=1\rangle\equiv |2\rangle$. These states possess
scattering properties that offer a natural way to suppress the
phase diffusion introduced by the $\hat{J}_z^{\,2}$ evolution;
namely, the $s$-wave scattering lengths for the processes
$|1\rangle|1\rangle \rightarrow |1\rangle|1\rangle$,
$|1\rangle|2\rangle \rightarrow |1\rangle|2\rangle$, and
$|2\rangle|2\rangle \rightarrow |2\rangle|2\rangle$,
respectively, are $a_{11}=100.40 a_0$, $a_{12}=97.66 a_0$, and
$a_{22}=95.00 a_0$~\cite{mertes07a}, with $a_0$ being the Bohr
radius, which implies that $\gamma_2\simeq0$. Consequently, the $\hat{J}_z^{\,2}$ term becomes negligible, and
the effective dynamics is simply described by the $N \hat{J}_z$
coupling.  This $N \hat{J}_z$ coupling is that of a linear Ramsey interferometer (i.e., a coupling proportional to $\hat{J}_z$), which accumulates phase at a rate enhanced by a factor of $N\eta_N$. This allows the coupling constant
$\gamma_1$ to be estimated with a sensitivity that scales as
$1/N^{3/2}\eta_N$. Notice that the exact scaling can only be
determined once the trapping potential is specified, since the
$N$ dependence of $\eta_N$ depends on trap geometry.

\subsection{Nonlinear Ramsey interferometry}
\label{Ramsey}

As in typical Ramsey interferometry schemes, our protocol runs as
follows.  The atoms are first condensed to the state
$\psi_N(\vec{r})|1\rangle$, and a fast optical pulse suddenly creates
the superposition state $\psi_{N}(\vec{r})(|1\rangle +
|2\rangle)/\sqrt{2}$ for each atom. The atoms are then allowed to
evolve freely for a time~$t$, which brings the atomic state
to [$\psi_{N,1}(\vec{r},t)|1\rangle +
\psi_{N,2}(\vec{r},t)|2\rangle]/\sqrt{2}$. A second transition
between the hyperfine levels is then used to transform any coherence
between the two modes into population information that is finally
detected. For this second transition, we choose a $\pi/2$ rotation about
the Bloch $x$ axis, changing the atomic state to
\begin{equation}
    \label{eq:GPfinalstate}
    \frac{1}{2}\bigl(\psi_{N,1}-i\psi_{N,2}\bigr)|1\rangle-
    \frac{i}{2}\bigl(\psi_{N,1}+i\psi_{N,2}\bigr)|2\rangle\;.
\end{equation}
The detection probabilities for each
hyperfine level,
\begin{equation}
    \label{eq:probs}
    p_{1,2}=\frac{1}{2}\bigl[1\mp
    \mbox{Im}(\langle\psi_{N,2}|\psi_{N,1}\rangle)\bigr],
\end{equation}
are modulated by the overlap of the two spatial wave functions,
\begin{equation}
\langle\psi_{N,2}|\psi_{N,1}\rangle=
\int d^3 r\,\psi_{N,2}^*\psi_{N,1}\;.
\label{ponetwo}
\end{equation}
This implements a measurement of
$\hat{J}_y$.

Within the Josephson approximation of Eq.~(\ref{BEChamil2}), the only
effect of the evolution is to introduce a differential phase shift
$N\eta_N\gamma_1 t/\hbar$ between the two modes. This implies that
ideally the probabilities oscillate as
\begin{equation}
\label{idealsignal}
 p_{1,2}=\frac{1}{2}\bigl[1\mp\sin(\Omega_N t)\bigr],
\end{equation}
where
\begin{equation}
\Omega_N\equiv N\eta_N\gamma_1/\hbar
\label{OmegaN}
\end{equation}
is the idealized fringe frequency.  This fringe pattern allows one to
estimate the coupling constant $\gamma_1$ with an uncertainty given
by
\begin{equation}
 \delta\gamma_1 =
 \frac{\langle(\Delta\hat{J}_y)^2\rangle^{1/2}}{|d\langle\hat{J}_y\rangle/d\gamma_1|}
 \sim \frac{1}{\sqrt{N}N\eta_N}\;.
\end{equation}

\section{Numerical simulations}
\label{numerics}

The several simplifying assumptions in the proposed model make it
straightforward to see how a scaling approaching $1/N^{3/2}$ can be
obtained. Those assumptions were made based on rough calculations
that suggest that the protocol is implementable with current
techniques~\cite{boixo09}. The main purpose of this paper is to
examine the validity of those assumptions by numerically simulating
the discussed interferometry scheme under realistic experimental
conditions.

\subsection{Spreading of the BEC wave function}
\label{spreading}

As emphasized before, the exact scaling of the detection sensitivity
ultimately depends on how $\eta_N$ varies with the number of atoms in
the condensate, which is essentially determined by the geometry of
the trapping potential, considering that $\eta_N^{-1}$ is a measure
of the volume occupied by the condensate. Because of the
repulsive interactions, the expansion of the BEC with increasing $N$
dilutes the effective nonlinear couplings, which can ruin the
enhanced scaling of the sensitivity~\cite{boixo09}.  The expansion of
the atomic cloud can be reduced by using a potential with hard walls,
which suppresses the $N$ dependence of $\eta_N$.  Another strategy is
to operate in traps of effectively lower dimension so that the
condensate wave function has fewer dimensions to spread into.

We thus determine the effect of the spreading of the condensate
wave function captured by $\eta_N$ by numerically integrating
the time-independent, three-dimensional GP
equation~(\ref{timeindepGPE})~\cite{dion07}. We restrict our
numerical analysis to the case of highly elongated BECs, which
according to previous results~\cite{boixo09}, offer the best
scalings. We assume that the BEC is tightly confined by a transverse
harmonic potential and loosely trapped by a power-law potential;
i.e., we consider cylindrically symmetric trapping potentials of
the form
\begin{equation}
\label{potential}
    V(\rho,z) = \frac{1}{2}(m\omega_T^2 \rho^2 + k z^q),
\end{equation}
with $q = 2, 4,$ and $10$.  These three potentials allow us to
explore how the results depend on the hardness of the potentials.
Notice that the limit $q\rightarrow \infty$ recovers the case of a
hard-walled trap.

In the so-called quasi-one-dimensional (quasi-1D) regime, the scattering interaction does not
drive any appreciable dynamics in the transverse directions.  One can
thus approximate the condensate wave function by the product ansatz
\begin{equation}
\label{ansatz}
    \psi_N(\rho,z) = \chi_0(\rho)\phi_N(z),
\end{equation}
where $\chi_0$ is the ground-state wave function of the transverse
harmonic potential and $\phi_N$ is the solution of the
one-dimensional, longitudinal GP equation
\begin{equation}
\label{1DGPE}
\left(-\frac{\hbar^{2}}{2 m}\frac{d^2}{dz^2}+
\frac{1}{2} k z^q + g_{11}N\eta_{T}|\phi_N|^{2}\right)\phi_N = \mu_L \phi_N\;.
\end{equation}
Here $\mu_L=\mu_N-\hbar\omega_T$ is the longitudinal part of the
chemical potential and $\eta_T$ is the inverse transverse
cross section of the condensate, given by
\begin{equation}
    \label{etaT}
    \eta_T = \int d^2\rho\,|\chi_0(\rho)|^4 = \frac{1}{2\pi\rho_0^2},
\end{equation}
where $\rho_0=\sqrt{\hbar/m\omega_T}$.

This quasi-1D approximation is valid only as long as the number of
atoms in the condensate is small compared to an (upper) critical atom
number $\bar{N}_T$, which is specified by determining when the
scattering energy becomes as large as the transverse kinetic energy.
This condition sets the characteristic energy required to excite any
dynamics in the transverse dimensions.  We thus define $\bar{N}_T$ by
solving the equation
\begin{equation}
     \frac{g_{11}}{2} N \eta_N =
     \frac{\hbar^2}{2m}\int d^2\rho\,|\nabla\chi_0|^2\;.
\end{equation}
Generally this equation would have to be solved numerically, but for
$N=\bar{N}_T$, the kinetic energy term in Eq.~(\ref{1DGPE}) can be
neglected, and the longitudinal wave function is well approximated by
the Thomas-Fermi solution~\cite{leggett01a}
\begin{equation}
 \label{thomasfermi1d}
 |\phi_N(z)|^2 = \frac{\mu_{L}-k z^{q}/2}{N g_{11}\eta_{T}},
\end{equation}
where $\mu_L\equiv k z_N^q/2$ is determined from the normalization
condition for $\phi_N$.  This defines the Thomas-Fermi longitudinal
size of the trap to be
\begin{equation}
    z_N = \left(\frac{q+1}{q}\frac{Ng_{11}\eta_T}{k}\right)^{1/(q+1)}.
    \label{zN}
\end{equation}

Given the Thomas-Fermi approximate solution to the 1D GP equation~(\ref{1DGPE}), $\eta_N$ can be easily calculated from the condensate wave function~(\ref{ansatz}) and is found to be given by
\begin{equation}
\label{eta1d}
 \eta_N =
 \frac{q}{2q+1}\left(\frac{q+1}{q}\right)^{q/(q+1)}
 \left(\frac{k}{N g_{11}}\right)^{1/(q+1)}
 \left(\frac{1}{2\pi\rho_0^2}\right)^{q/(q+1)},
\end{equation}
which yields
\begin{equation}
\label{NT}
 \bar{N}_T=
 \frac{q}{2(q+1)}\left(\frac{2q+1}{q}\right)^{(q+1)/q}
 \frac{z_0}{a}\left(\frac{z_0}{\rho_0}\right)^{2/q}
 =\frac{q}{2(q+1)}\left(\frac{2q+1}{q}\right)^{(q+1)/q}
 N_T,
\end{equation}
where $z_0=(\hbar^2/mk)^{1/(q+2)}$ is an approximation to the bare
ground-state width in the longitudinal direction.

For the analysis presented in~\cite{boixo09}, it was not
necessary to keep track of the purely $q$-dependent coefficient
that appears in Eq.~(\ref{NT}), which was thus omitted from the
definition of the critical atom number $N_T$.  This coefficient
decreases from 1.3 for a harmonic trap and goes to 1 in the
limit of a hard trap ($q\rightarrow\infty$). In the numerical
analysis that we present here, we find that Eq.~(\ref{NT})
provides a better estimate of the critical atom number that
characterizes the crossover between the one- and
three-dimensional regimes, therefore justifying the change in
definition from $N_T$ to $\bar{N}_T$.

The analysis in~\cite{boixo09} introduced another (lower) critical
atom number $N_L$ as the number of atoms at which the longitudinal
kinetic energy is equal to the scattering energy.  The
one-dimensional Thomas-Fermi approximation~(\ref{thomasfermi1d}) is
only justified for atom numbers well above $N_L$.  For the potentials
and parameters we consider here, $N_L$ is less than ten atoms.

For the numerical integration, we set the transverse frequency
to 350 Hz and the longitudinal frequency to 3.5 Hz for the
harmonic case ($q=2$), with the result that
$\bar{N}_T\simeq14\,000$ atoms.  To compare the simulations for
the different power-law potentials, we choose the stiffness
parameter $k$ so that $\bar{N}_T$ remains the same for the two
other values of $q$; thus all the traps have the same
one-dimensional regime of atom numbers. For such choice of
parameters, we find $\rho_0\simeq 0.6\,\mu$m and the aspect
ratio of the traps ($\rho_0$:$z_0$) to be approximately equal to
1:10, 1:24, and 1:57, respectively, for $q=2,4,10$.  In
addition, according to Eqs.~(\ref{zN}) and~(\ref{NT}), when
$N=\bar{N}_T$, the condensate aspect ratios ($\rho_0$:$z_N$) are
1:158, 1:146, and 1:138 (for $q=2, 4, 10$). These parameters are
typical of those in elongated BECs~\cite{jo07}.

\begin{figure}[ht]
\includegraphics[scale=.75]{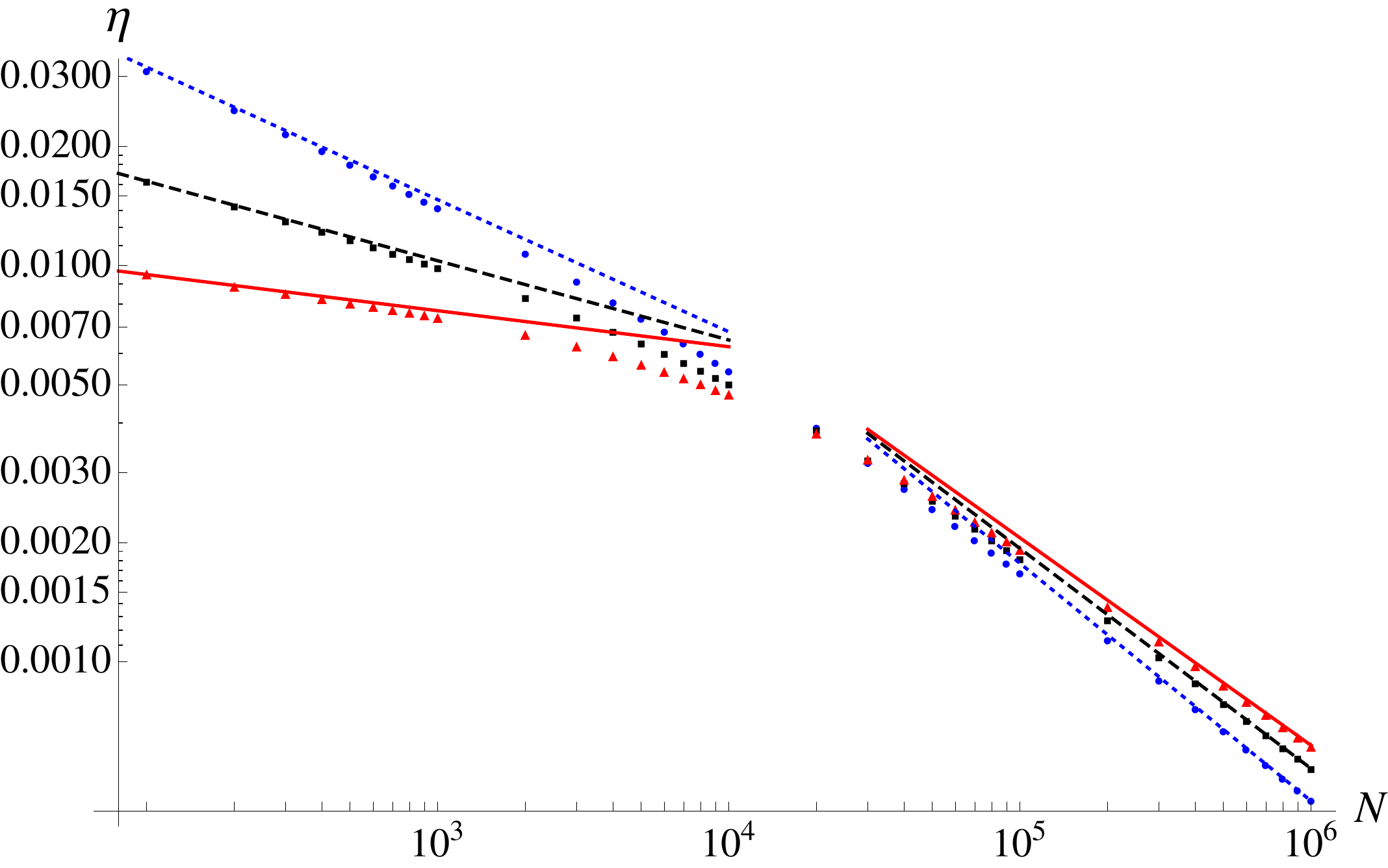}
\caption{(Color online) $N$ dependence of the inverse volume $\eta_N$
in harmonic trap units.  The points correspond to the results of the
numerical integration of the 3D GP ground-state solution for
different trap geometries: circles (blue) correspond to $q=2$,
squares (black) to $q=4$, and triangles (red) to $q=10$. The
respective Thomas-Fermi predictions for the 1D
[Eq.~(\protect\ref{eta1d})] and 3D regimes are the dotted (blue) line
for $q=2$, dashed (black) line for $q=4$, and solid (red) line for
$q=10$~\protect\cite{labeling}.  The stiffness parameter $k$ of the
trapping potential is chosen so that the crossover from 1D to 3D
behavior occurs at $\bar N_T\simeq14\,000$ for all three values of
$q$.}\label{etaplot}
\end{figure}

We numerically compute $\eta_N$ for the trapping
potentials~(\ref{potential}) and different atom numbers by first
solving the time-independent, three-dimensional GP
equation~(\ref{timeindepGPE})~\cite{dion07}. In Fig.~\ref{etaplot} we
plot the numerical results for $\eta_N$ as a function of the number
of atoms in the condensate for the three different values of
$q$ and compare the numerical results with the
Thomas-Fermi approximation in both the 1D and 3D regimes. These
results clearly show how the spreading of the condensate wave
function with increasing $N$ affects the scaling. For $N\ll
\bar{N}_T$, the tight transverse trap prohibits the spreading in the
radial direction.  The BEC can only expand in the longitudinal
dimension.  Such one-dimensional behavior is well described by
Eq.~(\ref{eta1d}), which predicts the scaling $N^{-1/(q+1)}$.  As $N$
approaches $\bar{N}_T$, the atomic repulsion gets stronger than the
radial confinement, and one sees deviations from the quasi-1D
behavior. Although the full expansion in the crossover regime can
only be determined numerically, we find that for $N
\lesssim\bar{N}_T$, one can predict the correct spreading with $N$ by
means of perturbative techniques, which will be presented in a
forthcoming publication.  For $N\gg \bar{N}_T$, the scattering term
in Eq.~(\ref{timeindepGPE}) becomes dominant, and the transverse
potential can no longer suppress the growth in the radial dimension.
In fact, the BEC enters the full three-dimensional Thomas-Fermi
regime, in which the entire kinetic energy becomes negligible, and it
is found that $\eta_N$ varies as $N^{-(q+1)/(2q+1)}$~\cite{boixo09}.

From the numerical evaluation of $\eta_N$, it is straightforward to
determine the exact scaling of $\delta \gamma_1$ with the number of
atoms in the condensate and thus verify that in all cases it is
better than $1/N$.

\subsection{Ramsey fringes}
\label{dynamics}

The dynamics of the two-mode $^{87}$Rb BEC we consider here is well
described by a mean-field approach which neglects the entanglement
and associated phase diffusion generated by the $\hat{J}_z^{\,2}$
interaction~\cite{mertes07a,anderson09}. In this approximation, the
wave functions for the two hyperfine levels evolve according to the
time-dependent, coupled, two-mode GP equations
\begin{equation}
\label{coupledGP}
i\hbar \frac{\partial\psi_{N,\alpha}}{\partial t} =
        \Biggl(-\frac{\hbar^2}{2m}\nabla^2+V+
        \sum_{\beta=1}^2g_{\alpha\beta}N_\beta|\psi_{N,\beta}|^2
        \Biggr)
        \psi_{N,\alpha}, \quad \alpha =1,2,
\end{equation}
which take into account the effect of the different scattering
processes on the evolution of each mode wave function~\cite{note1}.
Here $N_1$ and $N_2$ denote the respective (mean) populations of
levels~1 and~2.

We simulate the Ramsey interferometry scheme presented in
Sec.~\ref{Ramsey} as follows. Assuming that the atoms can be
prepared in the superposition state $\psi_{N}(\vec{r})(|1\rangle
+ |2\rangle)/\sqrt{2}$ with unit fidelity, we first integrate
Eq.~(\ref{timeindepGPE}) to find $\psi_N(\vec{r})$ and then
evolve it for a time~$t$ according to the coupled,
three-dimensional GP equations~(\ref{coupledGP}) for the
different potentials~(\ref{potential})\cite{xmds}. Finally,
supposing that the detection procedure can be considered
instantaneous, we find the spatial overlap of the computed
two-mode wave functions, $\psi_{N,1}(\vec{r},t)$ and
$\psi_{N,2}(\vec{r},t)$, which modulates the detection
probabilities $p_{1,2}(t)$ of Eq.~(\ref{ponetwo})~\cite{note2}.

\begin{figure}
\includegraphics[scale=.75]{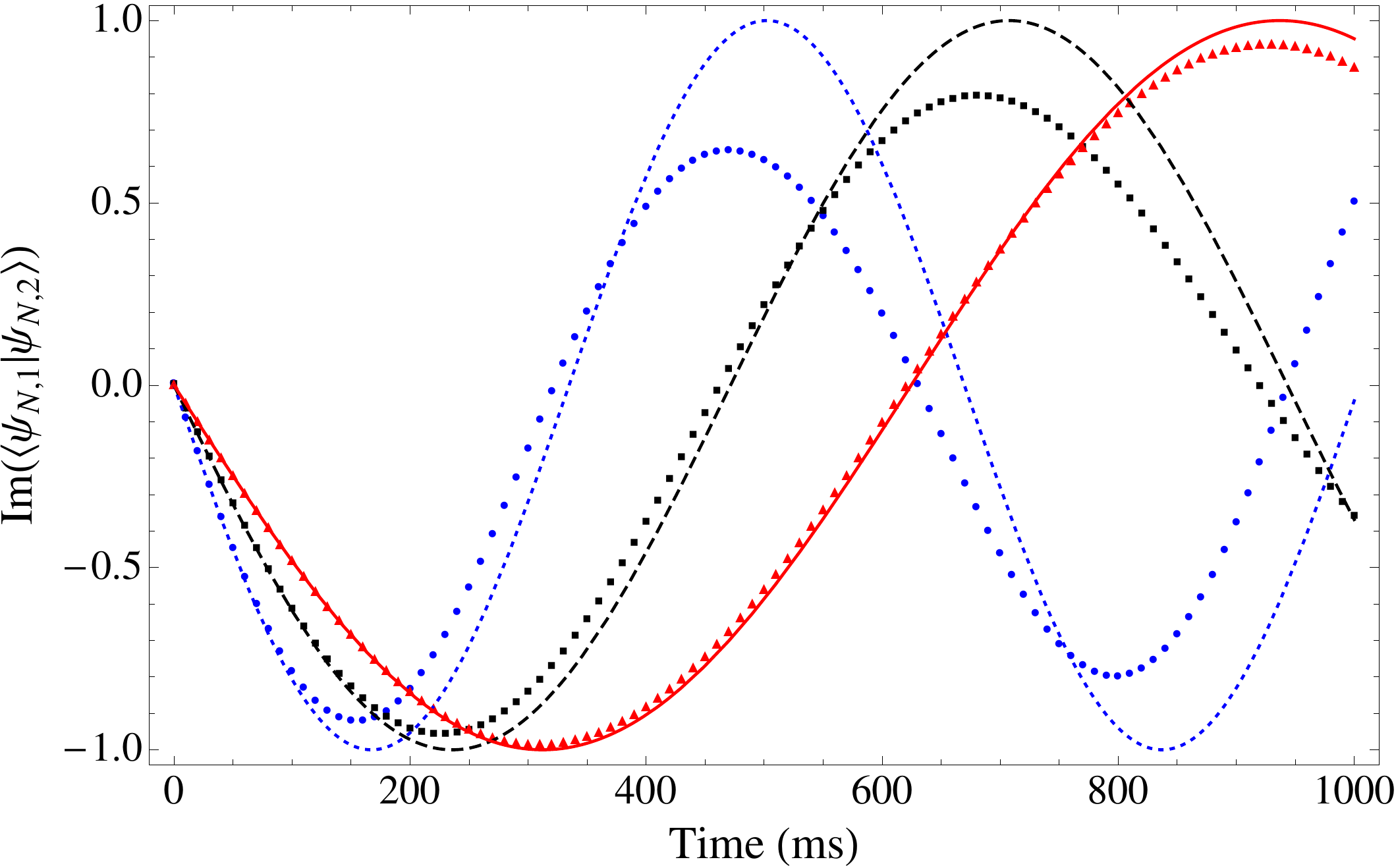}
\caption{(Color online) Ramsey fringes for a cigar-shaped $^{87}$Rb
BEC of 1\,000 atoms (labeling convention as in
Fig.~\protect\ref{etaplot}~\cite{labeling}). The points represent the
numerical results of the integration of the coupled, two-mode,
three-dimensional GP equations~(\protect\ref{coupledGP}) for the
different trapping potentials~(\protect\ref{potential}), whereas the
lines are the respective idealized Josephson-approximation
predictions~(\protect\ref{idealsignal}), with the value of $\eta_N$
supplied by the numerics of Sec.~\ref{spreading}.  The Josephson
approximation improves as the trap gets harder.}
\label{overlap1000}
\end{figure}

Figure~\ref{overlap1000} shows the resulting Ramsey fringe pattern
for a BEC of 1\,000 atoms, as well as the idealized,
Josephson-approximation fringe pattern~(\ref{idealsignal}), in which
we use the numerical value of $\eta_N$ found for the
potentials~(\ref{potential}), as described in Sec.~\ref{spreading}.
The agreement between the idealized fringe pattern and the numerical
results for the tenth-order potential is quite remarkable in view of
the complete neglect of spatial evolution by the Josephson
approximation. As $q$ decreases or time increases, however, the
simulated fringe pattern clearly deviates from the simplified
dynamics described by the Josephson approximation.  Such discrepancy
reveals, in fact, the breakdown of the Josephson approximation.

\begin{figure}
\includegraphics[scale=.75]{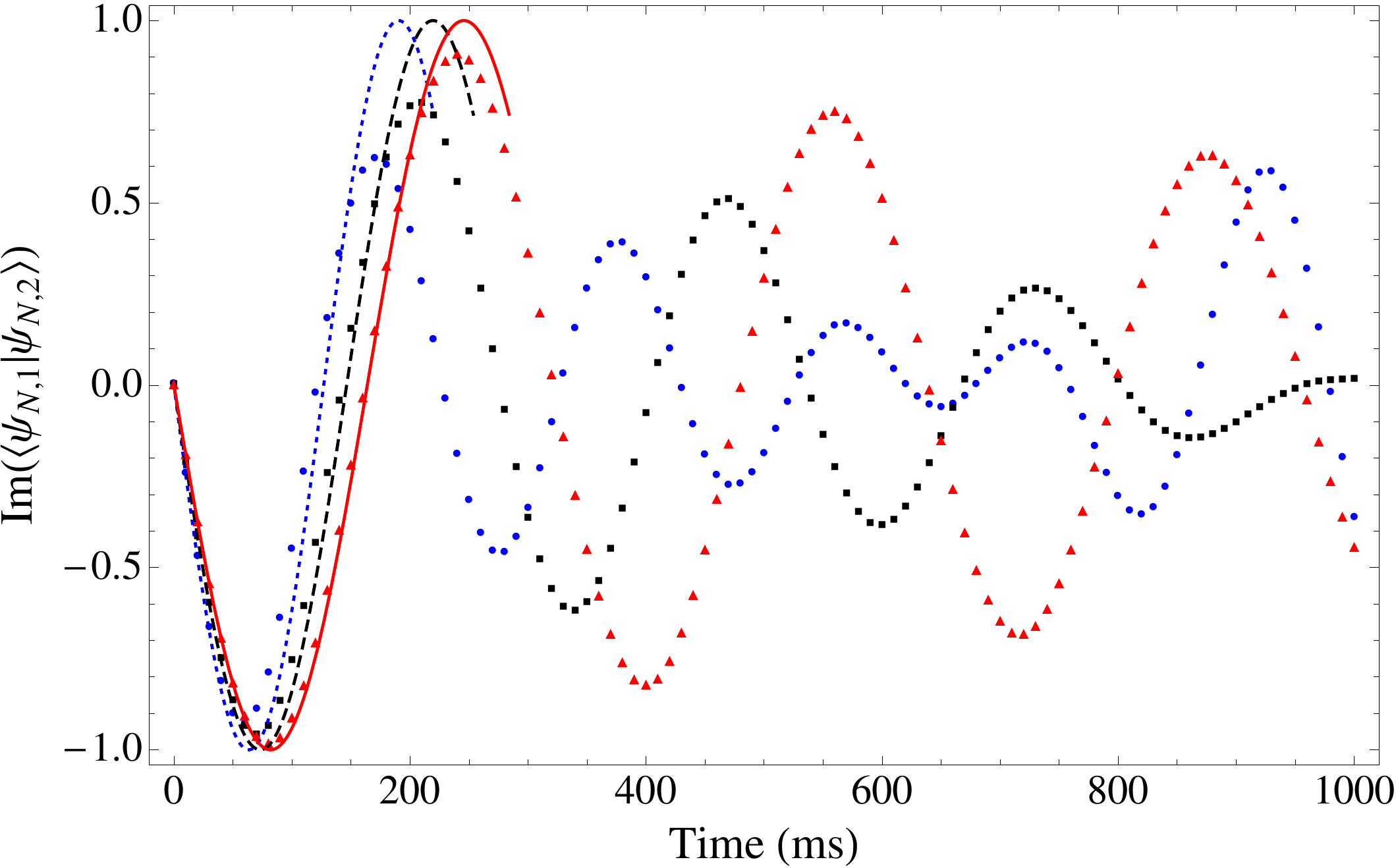}
\caption{(Color online) Ramsey fringes for a cigar-shaped $^{87}$Rb
BEC of 5\,000 atoms (labeling convention as in
Fig.~\protect\ref{etaplot}~\cite{labeling}). The points represent the
numerical results of the integration of the coupled, two-mode,
three-dimensional GP equations~(\protect\ref{coupledGP}) for the
different trapping potentials~(\protect\ref{potential}), whereas the
lines are the respective idealized, Josephson-approximation
predictions~(\protect\ref{idealsignal}).  Here we only plot the
idealized fringe pattern~(\protect\ref{idealsignal}) for short times,
since it quickly deviates from the simulated nonlinear evolution; the
deviation is a consequence of the differentiation of the wave functions
of two modes as they evolve separately under the coupled GP
equations.  (Figure~\protect\ref{closeup} shows a closeup of the
first 120\,ms.)} \label{overlap5000}
\end{figure}

\begin{figure}
\includegraphics[scale=.75]{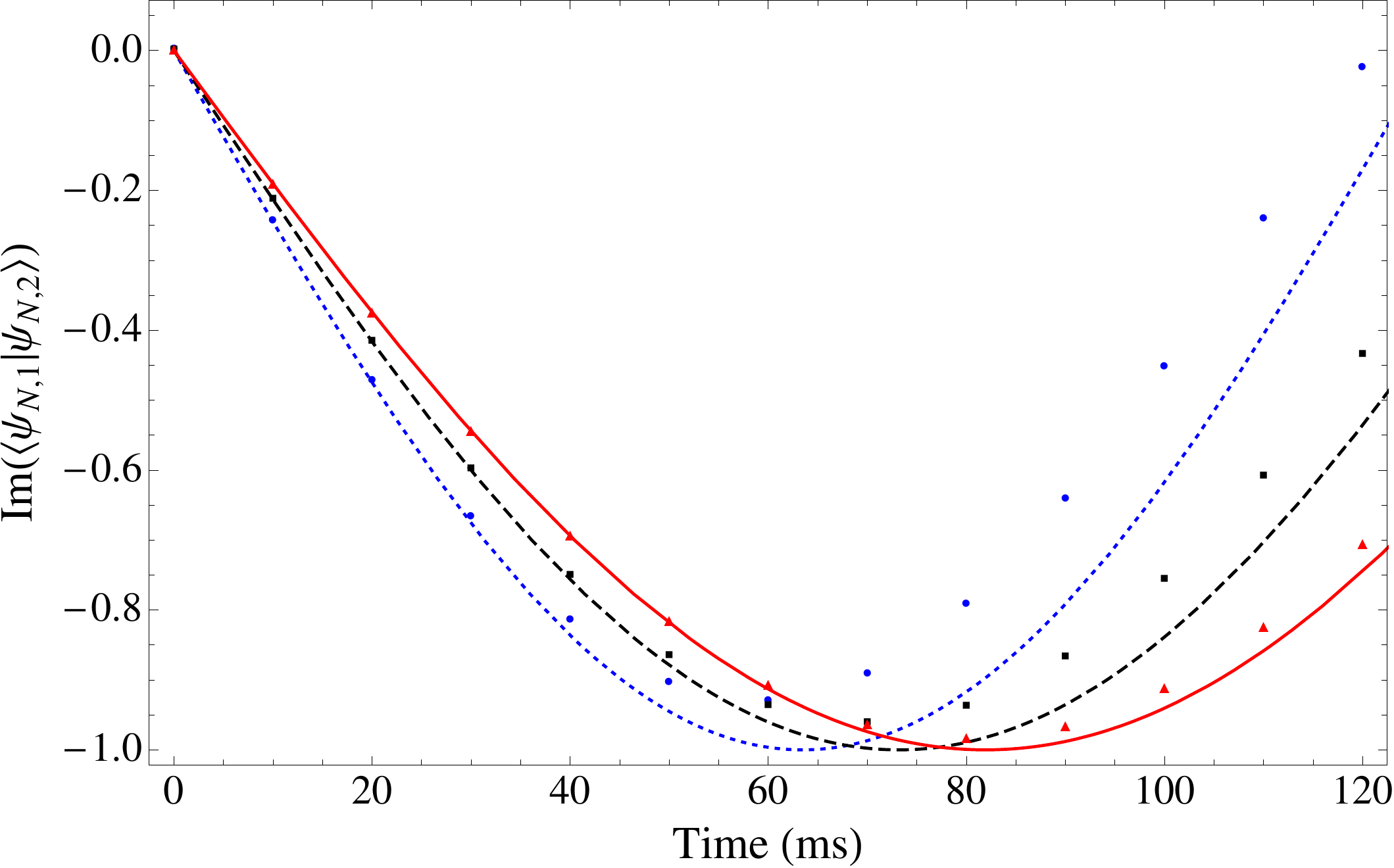}
\caption{(Color online) Closeup of Fig.~\protect\ref{overlap5000}
for the first 120 ms.  The Josephson approximation improves as the
trap gets harder.}
\label{closeup}
\end{figure}

The breakdown of the Josephson approximation becomes more
evident in the case of 5\,000 atoms shown in
Fig.~\ref{overlap5000}. Due to the difference in the scattering
lengths and the scattering potentials, each wave function has a
complex nonlinear evolution that, except for short times, can no
longer be approximated as no evolution at all, as is assumed by
the Josephson approximation.  The short-time behavior can be
better seen in Fig.~\ref{closeup}, where we plot the Ramsey
fringes for up to 120\,ms.  For longer times, the wave functions
differentiate spatially, which leads to the reduction in fringe
visibility and the change in the fringe frequency seen in
Figs.~\ref{overlap1000} and~\ref{overlap5000}. For $q=2$ and
5\,000 atoms, the fringe pattern is already entering a revival
before 1\,s. The fringe visibility is clearly better preserved
by going to a harder trap.

It is worth emphasizing that our results indicate that the integrated
phase shift that we are interested in detecting is accumulated more
rapidly than the time scale for the two wave functions to
differentiate spatially.  Moreover, for the regimes we consider and
in view of typical Ramsey pulses and detection times ($\lesssim
1\,$ms)~\cite{matthews98a,mertes07a,anderson09}, our simulations
confirm that the Josephson model holds for a length of time that is
sufficient to implement the metrology scheme.

In the following section, we analyze an alternative analytical model,
proposed in~\cite{boixo09}, which attempts to provide a better
description of the nonlinear detection signal shown in
Figs.~\ref{overlap1000} and~\ref{overlap5000}, by allowing the wave
functions to accumulate a position-dependent phase shift.

\section{Differentiation of the spatial wave functions}
\label{improvedmodel}

As already pointed out in the previous section, the distinct
scattering lengths of the allowed $s$-wave collisions for the
$^{87}$Rb atoms ultimately make each wave function evolve differently
in a nontrivial way~\cite{colson78}. In fact, it is known that due to
the interspecies repulsion, the two modes tend to separate
spatially~\cite{matthews98a}.  Before the modes segregate, however,
the effect of the different nonlinearities is to produce a relative
phase between the two modes that depends on the local density within
the condensate.

All these phenomena have recently been observed in a ground-breaking
set of experiments. In~\cite{anderson09}, Anderson~{\em et~al.}
measured the position-dependent phase shifts in the same two-mode
$^{87}$Rb BEC that we consider here, whereas Mertes~{\em
et~al.}~\cite{mertes07a} demonstrated the nonequilibrium separation
dynamics of the binary superfluid.  The details of both experiments
were shown to be well accounted for by numerical integrations of the
coupled, two-mode GP equations~(\ref{coupledGP}), with an additional
phenomenological loss term included.

In short, the two-mode dynamics can be explained as follows. For
short times, the integrated part of the relative phase, which
corresponds to the average difference in the energies of the
scattering processes, is the dominant dynamical effect and provides
the signal for our measurement protocol.  The residual
position-dependent part of the relative phase affects the two-mode
dynamics on a somewhat longer time scale and reduces the visibility
of the interference fringes on which the detected signal relies.
Eventually, the position-dependent phases drive differences between
the atomic densities associated with the two hyperfine levels, and
this leads to spatial separation of the two modes.

The above-described effects occur on different time scales,
which were estimated in~\cite{boixo09} to be sufficiently
different that the metrology protocol could be successfully
implemented.  The analytical estimates suggest that making the
longitudinal trap harder leads to a greater separation of these
three time scales.  In order to retain good fringe visibility,
the required operation time scale of the protocol was estimated
to be well within the first fringe, which we can confirm from
our simulations and is illustrated in Figs.~\ref{overlap1000}
and~\ref{overlap5000}.

In an attempt to model the complex dynamics of the two-mode
condensate, we modify our analytical description by allowing the
spatial wave functions to acquire a position-dependent phase shift in
addition to the uniform phase shift of the Josephson approximation.
Since the spatial segregation of the modes occurs on a longer time
scale than the accumulation of a position-dependent phase shift, it
is not relevant to this discussion.  Moreover, for the regimes that
concern us, we can include the position-dependent phase shift in a
quite straightforward way.  As before, we limit our analysis to the
trapping potentials~(\ref{potential}) and to quasi-1D BECs.  In
addition, considering our numerical simulations, we focus on the
particular case of both modes being equally populated, although a
more general discussion can be found in~\cite{boixo09}.

As in the case of the ground state of a single-mode BEC in the
one-dimensional regime, which was discussed in Sec.~\ref{spreading},
the wave functions of the two modes can be approximated by the
product of transverse and longitudinal wave functions,
\begin{equation}
\label{quasi1dapprox}
\psi_{N,\alpha}(\rho,z,t)=e^{-i\omega_T t}
\chi_0(\rho)\phi_{N,\alpha}(z,t), \quad \alpha =1,2,
\end{equation}
where $\chi_0$ is the previously defined time-independent, Gaussian
ground state in the transverse dimensions.  The longitudinal wave
functions satisfy the time-dependent, coupled, longitudinal GP
equations
\begin{equation}
\label{coupledGPlong}
i \hbar \frac{\partial\phi_{N,\alpha}}{\partial t} =
        \biggl(-\frac{\hbar^{2}}{2 m}\frac{d^2}{dz^2}+ \frac{1}{2} k z^q +
        \frac{1}{2}N\eta_T\sum_{\beta}g_{\alpha\beta}|\phi_{N,\beta}|^2\biggr)
        \phi_{N,\alpha},
\end{equation}
where we have used our assumption that $N_1=N_2=N/2$.

For the traps and atom numbers that we are considering, it is legitimate
to ignore the kinetic-energy terms in
Eq.~(\ref{coupledGPlong})~\cite{boixo09}, as is done in the 1D
Thomas-Fermi approximation.  Within this approximation, the
probability densities do not change with time, i.e.,
\begin{equation}
|\phi_{N,\alpha}(z,t)|^2=|\phi_N(z,0)|^2\equiv q_0(z),
\end{equation}
with $|\phi_N(z,0)|^2$ given by Eq.~(\ref{thomasfermi1d}); hence the
evolution under the coupled GP equations simply introduces a phase
that depends on the local atomic linear density,
\begin{equation}
    \label{eq:philong}
    \phi_{N,\alpha}(z, t) =
    \sqrt{q_0(z)}\exp\!\bigg[-\frac{it}{\hbar}\bigg(\frac{1}{2} k z^q +
    \frac{1}{2} N \eta_T q_0(z)
    \sum_{\beta} g_{\alpha \beta}\bigg)\bigg].
\end{equation}
This yields an overlap
\begin{equation}
\langle\psi_{N,2}|\psi_{N,1}\rangle
=\langle\phi_{N,2}|\phi_{N,1}\rangle
=\int dz\,q_0(z) e^{-i\delta\theta(z)},
\end{equation}
where the position-dependent differential phase shift is given by
\begin{equation}
\label{deltatheta}
\delta\theta(z)=\frac{q_0(z) N \eta_T \gamma_1\,t}{\hbar}=
\Omega_N t\left(1+\frac{q_0(z)-\eta_L}{\eta_L}\right)\;.
\end{equation}
Here
\begin{equation}
\eta_L\equiv\int dz\,q_0^2(z) =
     \frac{q}{2q+1}\left(\frac{q+1}{q}\right)^{q/(q+1)}
     \left(\frac{k}{N g_{11}}\right)^{1/(q+1)}
     \left(2\pi\rho_0^2\right)^{1/(q+1)},
\label{TFetaL}
\end{equation}
and we have separated out the integrated phase shift $\Omega_N t$,
whose angular frequency~(\ref{OmegaN}) is defined as in the Josephson
approximation.  This makes it clear that the residual
position-dependent phase shift adds a correction to the integrated
phase shift that we have previously calculated.

\begin{figure}[ht]
\includegraphics[scale=.75]{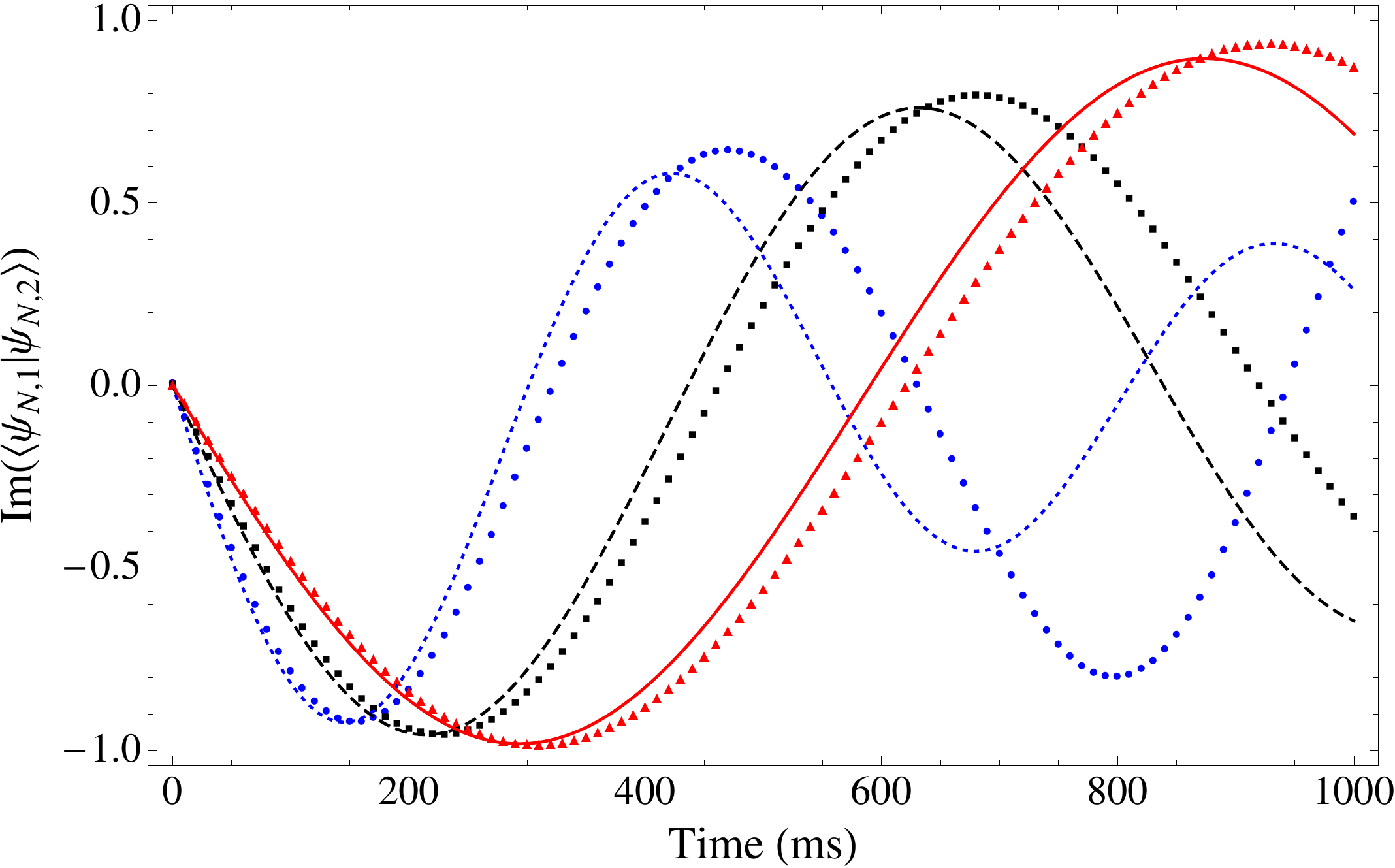}
\caption{(Color online) Same as Fig.~\protect\ref{overlap1000}
(1\,000 atoms), but here the lines correspond to the improved
analytical prediction~\protect(\ref{analyticaloverlap}) for the three
different values of $q$ (labeling convention as in
Fig.~\protect\ref{etaplot}~\cite{labeling}).  The improved model
succeeds in predicting a reduction in fringe visibility as time
increases, but it does not give a better estimate of the fringe
frequency.} \label{rawimproved1000}
\end{figure}

\begin{figure}[ht]
\includegraphics[scale=.75]{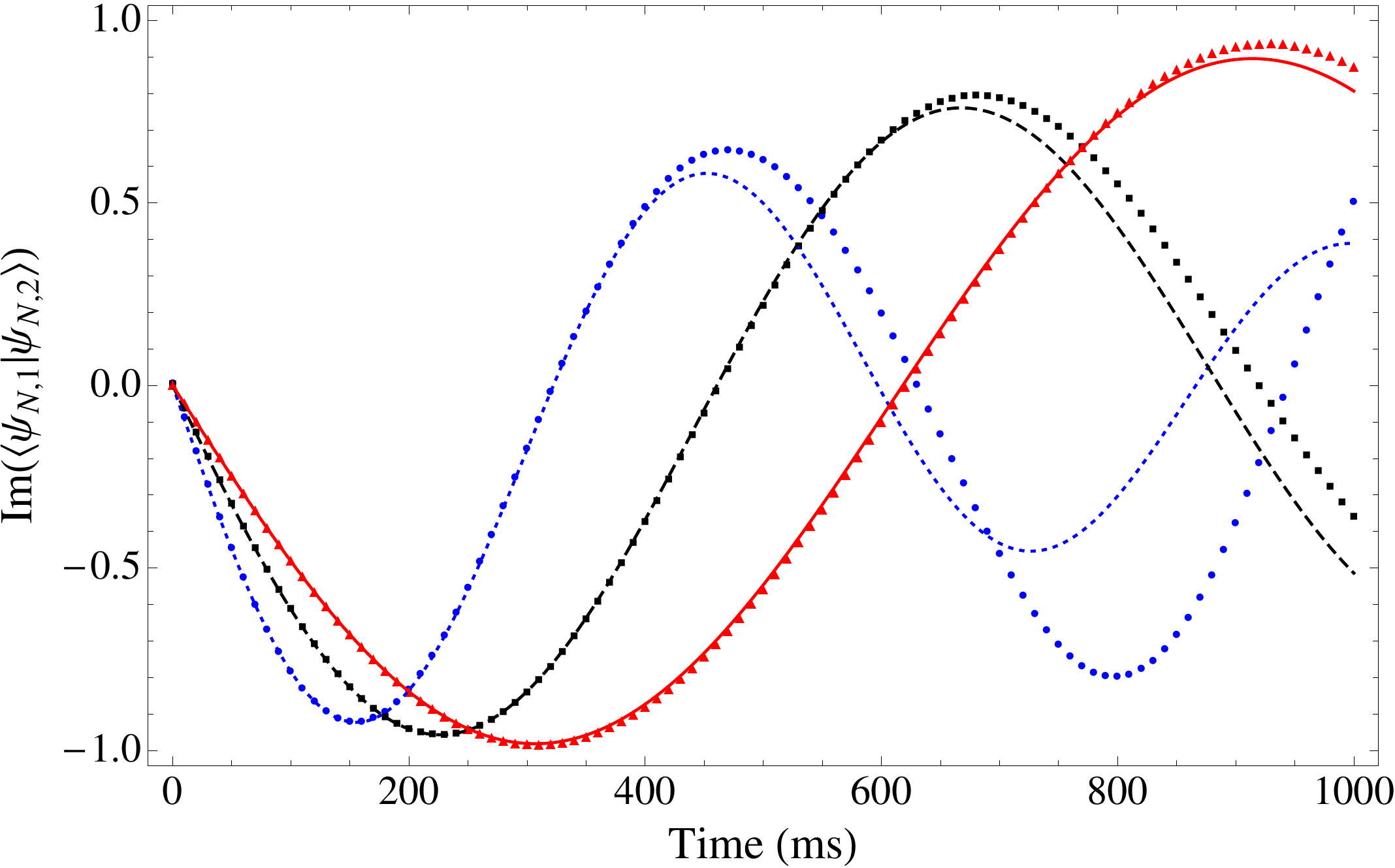}
\caption{(Color online) Same as Fig.~\protect\ref{overlap1000}
(1\,000 atoms), but here the lines correspond to the improved
analytical overlap~\protect(\ref{analyticaloverlap}), computed using
the numerically computed $\eta_N$ from Sec.~\protect\ref{spreading},
as described in the text, for the three different values of $q$
(labeling convention as in
Fig.~\protect\ref{etaplot}~\cite{labeling}).  The improved model,
with the {\it ad hoc\/} use of the numerically computed $\eta_N$,
provides a reasonably good account of the fringe frequency and of the
reduction in fringe visibility as time increases, with better
agreement for short times and for harder traps.} \label{improved1000}
\end{figure}

\begin{figure}[ht]
\includegraphics[scale=.75]{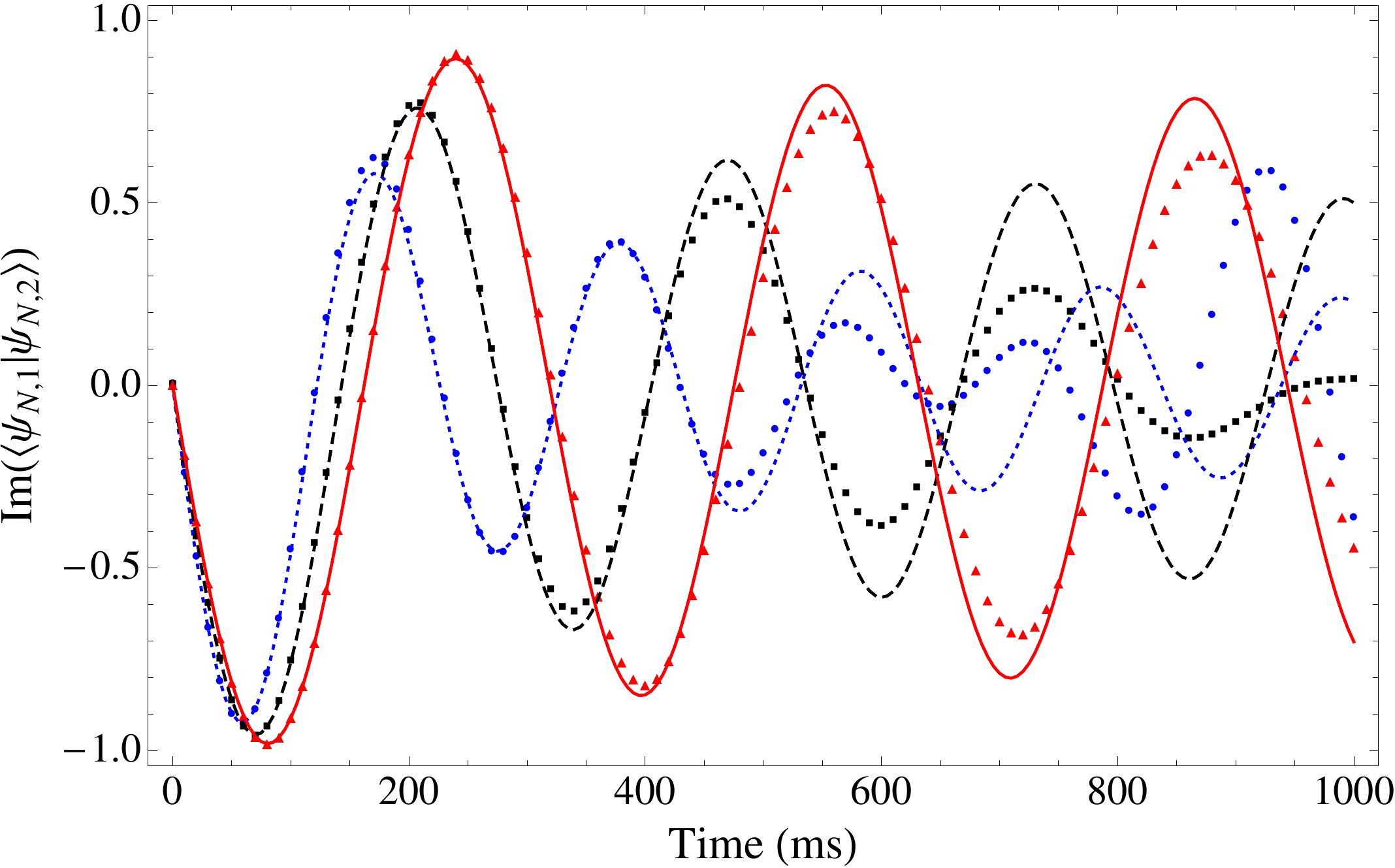}
\caption{(Color online) Same as Fig.~\protect\ref{overlap5000}, but
here the lines correspond to the improved analytical
overlap~(\protect\ref{analyticaloverlap}), computed using the
numerically computed $\eta_N$ from Sec.~\ref{spreading}, for the
three different values of $q$ (labeling convention as in
Fig.~\protect\ref{etaplot}~\cite{labeling}).  The improved model,
with the {\it ad hoc\/} use of the numerically computed $\eta_N$, is
surprisingly good even in the third fringe period for $q=10$.}
\label{improved5000}
\end{figure}

Putting all this together, we can write the overlap as
\begin{equation}
\label{analyticaloverlap}
\langle\psi_{N,2}|\psi_{N,1}\rangle= \int dz\,q_0 e^{-i q_0 N \eta_T \gamma_1\,t/\hbar}=
e^{-i\Omega_N t}\int dz\,q_0 e^{-i\Omega_N t(q_0-\eta_L)/\eta_L},
\end{equation}
whose imaginary part, as before, is responsible for the fringe
pattern in our interferometry scheme.  In
Fig.~\ref{rawimproved1000}, we compare the numerical fringes
with the imaginary part of the overlap~(\ref{analyticaloverlap})
for 1\,000 atoms.  We do exactly as the current approximation
instructs: we use the longitudinal Thomas-Fermi probability
density $q_0(z)$, the corresponding $\eta_L$ from
Eq.~(\ref{TFetaL}), and the transverse $\eta_T$ from the
transverse ground state $\chi_0$ [Eq.~(\ref{etaT})].  The
improved model captures an approximation to the reduction in
fringe visibility, with agreement with the numerics getting
better as $q$ increases, but it predicts a fringe frequency that
is too large. Indeed, by comparing Fig.~\ref{rawimproved1000} to
Fig.~\ref{overlap1000}, one sees that the frequency of the
improved model is too high by an amount that is somewhat larger
than the amount by which the Josephson approximation's frequency
is too low.

It is not hard to identify a source for this frequency
disparity.  The current approximation uses an atomic density
profile that comes from the product wave
function~(\ref{quasi1dapprox}), with the longitudinal wave
function coming from Eq.~(\ref{eq:philong}) and the transverse
wave function assumed to be the $N$-independent Gaussian ground
state of the transverse harmonic trap.  In contrast, the
frequency $\Omega_N$ we use in the Josephson-approximation plots
of Figs.~\ref{overlap1000} and~\ref{overlap5000} comes from
numerical computation of the 3D GP ground state.

We can test whether this is a source of the frequency disparity by
making an {\it ad hoc\/} adjustment to the model of this section.  In particular, in using the analytical overlap~(\ref{analyticaloverlap}), we can use the longitudinal Thomas-Fermi probability density $q_0(z)$ and its $\eta_L$ from Eq.~(\ref{TFetaL}), as the approximation instructs. We could instead adopt the {\it ad hoc\/} procedure of using the numerically computed $\eta_N$ plotted in Fig.~\ref{etaplot}; the transverse $\eta_T$ determined in this procedure from $\eta_T=\eta_N/\eta_L$ , is no longer that of the transverse ground state and acquires an $N$ dependence from $\eta_N$ and $\eta_L$.

In Figs.~\ref{improved1000} and~\ref{improved5000}, we compare
the numerical fringes with the imaginary part of the
overlap~(\ref{analyticaloverlap}), computed using the {\it ad
hoc\/} modification, for 1\,000 and 5\,000 atoms.  The improved
model, with this {\it ad hoc\/} adjustment, is surprisingly good
at predicting both the fringe frequency and the reduction in
fringe visibility, especially for $q=10$ (the
Josephson-approximation fringes would have the same frequency
discrepancy had we used the 1D Thomas-Fermi approximation and
the transverse ground state to determine $\eta_N$, instead of
using the numerical value).  We emphasize that the fringe
visibility is preserved better by going to harder traps. Within
the improved model, it is clear that the better fringe
visibility of harder traps is due to the fact that as $q$
increases, the trapping potential becomes more flat-bottomed,
making the atomic density profile more uniform across the trap
and and thus reducing the size of the residual
position-dependent phase shift.

It is clear that our improved analytical model does indeed
provide a more accurate description of the nonlinear evolution
of the two-mode condensate and consequently of the fringe
pattern of our protocol.  Notice, however, that for longer times
effects that are not considered in this model, such as mode
segregation, become significant and, therefore, the Ramsey
fringes can no longer be described by
Eq.~(\ref{analyticaloverlap}).  As already noted, for $q=2$ and
5\,000 atoms, the nonlinear evolution is undergoing a revival
well before $t=1\,$s, an effect that cannot be described within
our model.

Throughout this analysis we consider the case of a one-dimensional
BEC whose ground-state wave function is supposed to be well
approximated by the product ansatz~(\ref{ansatz}). In this
approximation one assumes that the effect of the scattering
interaction on the transverse degrees of freedom of the gas can be
completely neglected.  As we discuss in Sec.~\ref{spreading}, this is
a good approximation as long as the number of atoms in the condensate
is small compared to the critical atom number $\bar{N}_T$.  In fact,
from Fig.~\ref{etaplot} one sees that as $N$ approaches $\bar{N}_T$,
the product wave function~(\ref{ansatz}) fails to give an accurate
estimate of the inverse volume $\eta_N$.

The main reason for this discrepancy is that as $N$ approaches the
critical atom number $\bar N_T$, the condensate begins to spread in
the transverse dimensions. Indeed, the analysis in this section shows
that we obtain a reasonably good account of the fringe signal by
including a position-dependent phase shift to describe the reduction
in fringe visibility and by allowing $\eta_T$ to change with $N$ as
dictated by the numerical 3D ground-state wave function, thus
reflecting the spreading of the condensate in the transverse
dimensions.

\section{Conclusion}
\label{conclusion}

We present in this paper a detailed numerical analysis of a
recent proposal of a Ramsey interferometry scheme that takes
advantage of the nonlinear scattering interactions in a two-mode
$^{87}$Rb Bose-Einstein condensate to achieve detection
sensitivities that scale better than the $1/N$ limit of linear
metrology~\cite{boixo08b}.  In view of current experimental
techniques and typical parameters, this scheme is a feasible
proof-of-principle experiment that in terms of sensitivity
scaling, can outperform Heisenberg-limited linear
interferometry. The proposed protocol does not rely on
complicated state preparation or measurement schemes nor on
entanglement generation to enhance the measurement sensitivity.

We first analyze here how the scaling is affected by the
expansion of the atomic cloud as a function of the number of
atoms in the condensate and of the geometry of the trap, by
considering the case of quasi-1D BECs trapped by different
potentials. Later we find the exact dependence of the atomic
density with the atom number $N$, solving numerically the
single-mode, three-dimensional GP equation. This allows us to
pin down the exact scaling and to verify that in all the
considered cases a scaling better than $1/N$ can be achieved.

In addition, we simulate the proposed interferometric scheme and the
corresponding measurement signal by solving the two-mode, coupled,
three-dimensional GP equations.  Our numerical results not only
confirm the theoretical predictions derived in~\cite{boixo09}, but
also show that the assumption that the two modes share the same
spatial wave function is justified for a length of time sufficient to
run the metrology scheme.  For longer times, it becomes evident that
the Josephson Hamiltonian~(\ref{BEChamil2}) is unable to handle the
full two-mode dynamics because it ignores entirely the spatial
evolution of the condensate wave functions.

To get a more accurate description of the fringe signal, one needs to
take into account the spatial differentiation of the wave functions
of the two modes.  We formulate an improved model that partially
describes the differentiation of the wave functions by including a
position-dependent phase shift across the condensate.  This model,
based on a one-dimensional Thomas-Fermi approximation, gives a
considerably refined account of the fringe signal of our protocol.

Our analysis shows that as the number of atoms in the condensate
approaches the critical atom number $\bar{N}_T$, deviations of the
assumed product wave function from the numerically computed initial
condensate wave function result in less accurate analytical estimates
of the oscillation frequency of the fringe pattern.  Preliminary
results indicate, however, that this effect can be described by means
of perturbative techniques, which treat the scattering interaction as
a perturbation to the single-particle transverse Hamiltonian and
which will be presented elsewhere. The perturbation theory indicates
that there should be corrections to the product wave function as well. For more hard-walled and
flat-bottomed potentials, we find that corrections to our idealized
models become less important, confirming that one should consider
potentials such as boxes or rings as the preferred architectures for
nonlinear BEC metrology.

\acknowledgments

This work was supported in part by the Office of Naval Research
(Grant No.~N00014-07-1-0304) and the National Science Foundation
(Grant No.~PHY-0903953 and No. PHY-0653596).  S.B. was supported by
the National Science Foundation under Grant No. PHY-0803371 through
the Institute for Quantum Information at the California
Institute of Technology.  A.D. was supported in part by the EPSRC
(Grant No. EP/H03031X/1) and the European Union Integrated Project
(QESSENCE).


\begin{thebibliography}{100}

\bibitem{giovannetti06a}
V. Giovannetti, S. Lloyd, and L. Maccone, Phys. Rev. Lett. {\bf 96},
010401 (2006).

\bibitem{luis04a}
A. Luis, Phys. Lett. A {\bf 329}, 8 (2004).

\bibitem{boixo08a}
S. Boixo, A. Datta, S. T. Flammia, A. Shaji, E. Bagan, and C. M.
Caves, Phys. Rev. A {\bf 77}, 012317 (2008).

\bibitem{boixo08b}
S. Boixo, A. Datta, M. J. Davis, S. T. Flammia, A. Shaji, and C. M.
Caves, Phys. Rev. Lett. {\bf 101}, 040403 (2008).

\bibitem{woolley08a}
M. J. Woolley, G. J. Milburn, and C. M. Caves, New J. Phys. {\bf 10},
125018 (2008).

\bibitem{boixo09}
S. Boixo, A. Datta, M. J. Davis, A. Shaji, A. B. Tacla, and C. M.
Caves, Phys. Rev. A {\bf 80}, 032103 (2009).

\bibitem{boixo07a}
S. Boixo, S. T. Flammia, C. M. Caves, and J. M. Geremia, Phys. Rev.
Lett. {\bf 98}, 090401 (2007).

\bibitem{rey07}
A. M. Rey, L. Jiang, and M. D. Lukin, Phys. Rev. A {\bf 76}, 053617 (2007).

\bibitem{choi08}
S. Choi and B. Sundaram, Phys. Rev. A {\bf 77}, 053613 (2008).

\bibitem{dalfovo99}
F. Dalfovo, S. Giorgini, L. P. Pitaevskii, and S. Stringari, Rev.
Mod. Phys. {\bf 71}, 463 (1999).
\bibitem{leggett01a}
A. J. Leggett, Rev. Mod. Phys. {\bf 73}, 307 (2001).

\bibitem{mertes07a}
K. M. Mertes, J. W. Merrill, R. Carretero-Gonzalez, D. J.
Frantzeskakis, P. G. Kevrekidis, and D. S. Hall, Phys. Rev. Lett.
{\bf 99}, 190402 (2007).

\bibitem{dion07}
C. M. Dion and E. Canc\`es, Comp. Phys. Comm. {\bf 177}, 787 (2007).

\bibitem{jo07}
G.-B.~Jo, J.-H. Choi, C.~A. Christensen, Y.-R. Lee, T.~A. Pasquini,
W.~Ketterle, and D.~E. Pritchard, Phys. Rev. Lett. {\bf 99},
240406 (2007).

\bibitem{labeling}
All the plots in this paper share the following labeling conventions:
points represent numerical data for the different trapping
potentials~(\ref {potential}), with circles (blue) corresponding to
$q=2$, squares (black) to $q=4$, and triangles (red) to $q=10$;
corresponding analytical expressions are plotted as lines: dotted
(blue) for $q=2$, dashed (black) for $q=4$ and solid (red) for
$q=10$.

\bibitem{anderson09}
R. P. Anderson, C. Ticknor, A. I. Sidorov, and B. V. Hall, Phys. Rev.
A {\bf 80}, 023603 (2009).

\bibitem{note1}
It is noteworthy that the coupled GP equations include a linear
approximation to $\hat {J}_z^{\,2}$, which corresponds to setting
$\hat {J}_z^{\,2}\simeq \langle \hat {J}_z\rangle ^2+2\langle \hat
{J}_z\rangle \Delta \hat {J}_z$, which should be valid for short
times.  Notice that because we consider the populations of the two
modes to be the same, it follows that $\langle \hat {J}_z\rangle =0$.

\bibitem{xmds}
To solve the coupled two-mode GP equations numerically, we use the
code generator {\it XmdS}, available at {\tt www.xmds.org}.

\bibitem{note2}
In our simulations we assume that one can neglect the evolution of
the spatial modes during the Ramsey pulses.  According to~\cite
{matthews98a,mertes07a,anderson09}, the two hyperfine levels of
$^{87}$Rb can be coupled by two-photon Rabi pulses whose duration is
of the order of 1 ms, which is much faster than the characteristic
time for the condensate to change its shape, thereby justifying such an
approximation.

\bibitem{matthews98a}
M. R. Matthews, D. S. Hall, D. S. Jin, J. R. Ensher, C. E. Wieman, E.
A. Cornell, F. Dalfovo, C. Minniti, and S. Stringari, Phys. Rev.
Lett. {\bf 81}, 243 (1998).

\bibitem{colson78}
W. B. Colson and A. L. Fetter, J. Low Temp. Phys. {\bf 33}, 231
(1978).

\end{thebibliography}
\end{document}